\documentclass[aps,prl,twocolumn,superscriptaddress,showpacs]{revtex4}
\usepackage{graphicx}
\usepackage{bm}

\begin{document}
\title{Demagnetized Electron Heating at Collisionless Shocks}
\author{David~Sundkvist}
\email{sundkvist@ssl.berkeley.edu}
\affiliation{Space Sciences Laboratory, University of California, Berkeley, CA. 94720 USA}
\author{F.~S.~Mozer}
\affiliation{Space Sciences Laboratory, University of California, Berkeley, CA. 94720 USA}
\date{November 14, 2012}

\begin{abstract}
Seventy measurements of electron heating at the Earth's quasi-perpendicular bow shock are analyzed in terms of Maxwellian-temperatures obtained from fits to the core electrons that separate thermal heating from supra-thermal acceleration. The perpendicular temperatures are both greater and lesser than expected for adiabatic compression. The average parallel and perpendicular heating is the same. These results are explained because, over the electron gyroradius, $\delta B/B\sim 1$ and $e\delta \phi/T_e\sim 1$, so electron trajectories are more random and chaotic than adiabatic. Because density fluctuations are also large, trapping and wave growth in density holes may be important.
\end{abstract}
\maketitle
Collisionless plasma shocks play an important role in astrophysical and solar system plasmas by redistributing supersonic flow energy into thermal
energy of plasma, fields, waves and accelerated particles. At the very heart of the physics of collisionless shocks is the question of dissipation, which is needed to limit the steepening of the shock \cite{kennel1985a,treumann2009a}. A shock steady state is reached when the nonlinear steepening is balanced by dissipation with the addition of dispersion. 

The shock heating mechanisms for the ions and electrons are believed to be very different. In constrast to the ions the problem of the electron heating is not fully understood, and remains controversial after some 40 years of research. Early theories focused on micro-instabilities such as ion-acoustic or lower-hybrid that should arise from the presence of unstable currents in the shock layer \cite{papadopoulos1985a}. 
The current consensus however is focusing on the presence of the macroscopic (DC) magnetic and electric fields and associated potential in the so called de Hoffman-Teller-frame. These fields inflate and also open up a hole in the electron population when crossing the shock, if the electrons are magnetized and effectively adiabatic \cite{goodrich1984a,scudder1986c,scudder1995a}. This process is reversible and additional scattering and/or instabilities are required to infill the distribution and thermalize the lower energy electrons.

Previous studies suggested that electrons conserve fluid adiabaticity ($T_{e,\perp}/B\sim \mathrm{constant}$, where $T_{e,\perp}$ is the electron temperature perpendicular to the magnetic field) across the shock \cite{scudder1986a,scudder1986b}, although a statistical study showed some heating above that expected from conservation of the adiabatic moment \cite{schwartz1988a}.
Early observations of super-adiabatic electron heating prompted the development of alternative heating mechanisms \cite{balikhin1993a,balikhin1994a,gedalin1995a,gedalin1995b,balikhin1998a} where electrons are non-adiabatic on a particle level and the electron trajectories are kinematically described. The demagnetized and non-adiabatic behaviour in these models is due to the presence of a large-scale (macroscopic) electric field gradient, resulting in super-adiabatic heating of electrons depending on the gradient scale. 
The shock transition scale size and other gradients are thus important not only for determining the importance of dispersive effects \cite{sundkvist2012a} in addition to dissipation \cite{schwartz2011a}, but also for determining the adiabaticity and resultant heating of the electrons. 
Non-adiabaticity of electrons has also been studied in simulations \cite{lembege2003a,savoini2005a,savoini2010a} were it was noted that different sub-populations of electrons could be over- or sub-adiabatic depending on injection angle. 

In this letter we use space plasma observations to show that the overall thermal electron heating is rarely strictly adiabatic but can be super- as well as sub-adiabatic (depending on Mach number), and that this is due to the fact that the ratio of perturbed to the average magnetic field, and the ratio of electric potential to thermal temperature (effectively the electric field gradient) are of the order of one, at all relevant length scales through the shock ramp transition. 
By introducing the concept of the perpendicular Maxwellian-temperature and contrasting it with both the conventional perpendicular moments-temperature and the adiabatic-temperature, we also show that previous studies have likely overestimated the amount of electron heating. We then discuss the properties of the perpendicular Maxwellian-temperature at 70 quasi-perpendicular bow shock crossings.

We use data from the THEMIS multi-spacecraft mission \cite{angelopoulos2008a}, 
during times when at least three of the THEMIS spacecraft were close
together, and the time between individual encounters of the Earth's bow
shock was short. 

As an example, we will consider the THEMIS A crossing on December 7, 2010, at 14:59:26 UT (Universal Time) near $X_{\mathrm{GSE}}=8.26$, $Y_{\mathrm{GSE}}=-7.53$ and $Z_{\mathrm{GSE}}=2.84$ (Geocentric Solar Ecliptic coordinates), within about 30 seconds of similar crossings by THEMIS D and THEMIS E. This was a quasi-perpendicular $(\theta_{Bn}=85^{\circ })$ shock crossing with super-critical Alfv\'enic and fast Mach numbers $(M_A=13, M_F=6)$.  $\theta_{Bn}$ is the angle between the upstream magnetic field and the normal to the shock and it was obtained from a minimum variance analysis of the magnetic field.  The shock normal direction in GSE was $\hat{\mathbf{n}} = (0.74,-0.67,-0.13)$.  Multi-spacecraft timing analysis gave a shock speed of 5.5 km/s, consistent with the above shock normal.  The magnetic field increased from 5 nT to 30 nT across the shock and the plasma density increased from 30 cm-3 to more than 300 cm-3.  Quarter-second snapshots of typical $75^{\circ}-105^{\circ}$ (hereafter called perpendicular) electron distributions across this bow shock are illustrated in Figure 1.  These distributions consist of low energy thermal cores and supra-thermal power law distributions of halo electrons, originating in the upstream solar wind. Maxwellian-temperatures are defined as the fits of Maxwell-Boltzmann distributions to the thermal cores of the plasma and are illustrated as the red curves in Figure 1.
\begin{figure}[tbp]
\includegraphics[width=0.9\columnwidth]{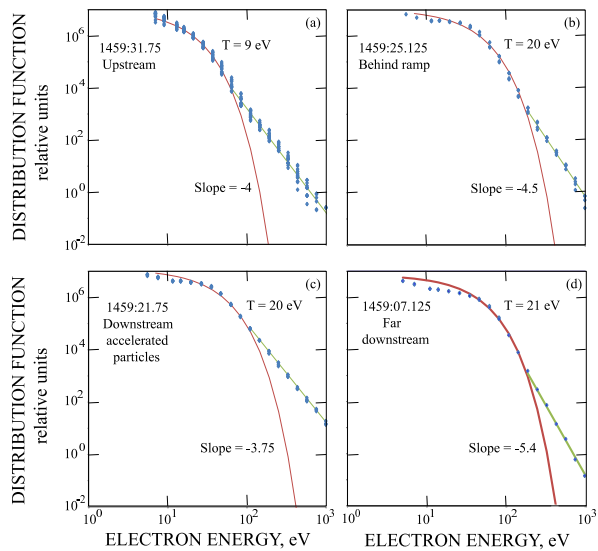}
\caption{Snapshots of the electron distribution function sampled over
0.25s, for pitch angles of $75^{\circ }-105^{\circ }$. The snapshots are
taken at distinctive regions across the shock. Arrows in 
Fig.~\ref{fig:ts} show where the snapshots were taken.
(a) Upstream region. (b) Behind the ramp. (c) Downstream. (d) Far downstream. }
\label{fig:composite}
\end{figure}
These and all distributions in this paper are analyzed in the spacecraft frame because their transformation to the more appropriate frame of the plasma has been shown to affect the Maxwellian-temperature estimate by no more than about one eV, because the plasma flow speed is small compared to the thermal electron speed.  It is also noted that the spectra have been adjusted for the spacecraft potential.  The four snapshots of the perpendicular electron distributions in Figure 1 are taken at select positions that highlight distinctive features of electron shock physics. 
The positions for these snapshots are indicated with arrows in Fig.~\ref{fig:ts}.
In the upstream region, panel (a), a mixture of 9 eV perpendicular core electrons co-exist with halo electrons of higher energy and, perhaps, some shock accelerated electrons.   Just downstream of the ramp in panel (b), the Maxwellian-temperature increased by a factor of two and flat-top distributions were seen. These electrons were observed with relatively little change in the thermal component further downstream, in panels (c) and (d).

Time-domain data across the shock of Figure~\ref{fig:composite} are illustrated in Figure~\ref{fig:ts}, in which the upstream direction is at the right of the figure and the shock ramp is indicated by the yellow rectangle.  The black dash-dotted curve in panel (a) is the perpendicular adiabatic-temperature expected for adiabatic compression of the plasma. It is computed from the magnetic field data by making the first invariant, $T_{e,\perp}/B = T_{e,\perp,sw}/B_{sw}$, a constant whose value was determined in the solar wind. The magnetic field is here sampled with 128 vectors/s by the fluxgate magnetometer. 
\begin{figure}[tbp]
\includegraphics[width=0.9\columnwidth]{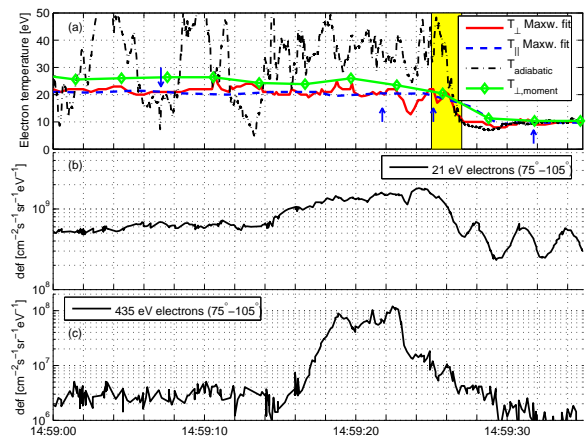}
\caption{Shock electrons sampled by the THEMIS A spacecraft. (a) Electron perpendicular and parallel temperatures. Fitting a Maxwellian to the thermal part of the electron distribution function gives a perpendicular (red solid) and parallel (blue dashed) heating of electrons that is less than that expected from adiabatic compression (black dashdot). Also plotted is the perpendicular moment-temperature that includes supra-thermal electrons, calculated using high resolution plasma data. Arrows indicate locations of distribution functions in Fig.~\ref{fig:composite}. (b) Differential energy flux (DEF) for 22 eV (thermal) perpendicular electrons (c) DEF 435 eV (non-thermal) perpendicular electrons.}
\label{fig:ts}
\end{figure}
The red solid curve in panel (a) is the perpendicular Maxwellian-temperature and the green curve with diamonds is the perpendicular moment-temperature obtained from moments of the spin period distribution function. The blue dashed curve is the parallel Maxwellian-temperature.
Because the moment-temperature includes contributions from the supra-thermal electrons, it is greater than the Maxwellian-temperature, as shown in panel (a).  In the current data set, the moment-temperature exceeds the Maxwellian-temperature by factors from 1.2 to 2.0.

Panels (b) and (c) of Figure 2 give the fluxes of 22 eV thermal electrons and 435 eV supra-thermal electrons.    The thermal electrons were heated predominately in the shock ramp and the more energetic electrons were accelerated about three seconds downstream of the ramp (as can also be seen in panel (c) of Figure 1).  Thus, thermal heating and supra-thermal electron acceleration arise from different physical processes, so separating them, as is done by considering the Maxwellian-temperature instead of the moments-temperature, is necessary for a more careful analysis of electron heating at the bow shock.
\begin{figure}[tbp]
\includegraphics[width=0.9\columnwidth]{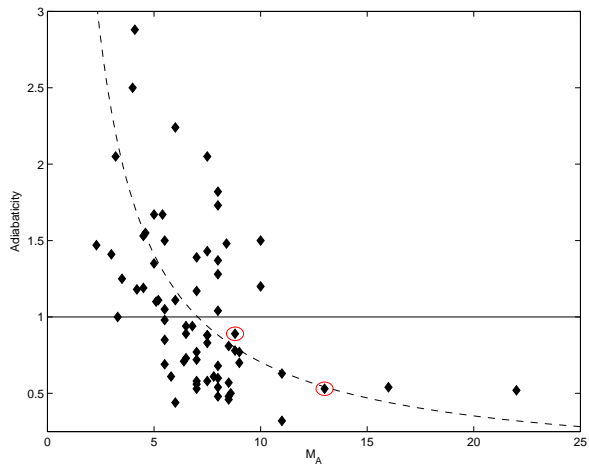}
\caption{Adiabicity $T_{\mathrm{Maxwellian}}/T_{\mathrm{adiabatic}}$ as a function of the Alfv\'en Mach number. The dashed curve is proportional to $1/M_A$. Circled points correspond to the two events presented in this Letter.}
\label{fig:adiabaticity}
\end{figure}
An important result in panel (a) of Figure 2 is that the thermal electron heating described by the Maxwellian-temperature is less than half that expected for adiabatic electron heating in the compressed magnetic field, as described by the adiabatic-temperature.  This is the first experimental evidence that electrons may be heated less than adiabatically at the bow shock. We note that in this event also the moment-temperature is less than that expected from adiabatic heating.

To determine statistically how thermal electrons are heated relative to adiabatic heating, 70 quasi-perpendicular bow shock crossings with $\theta_{Bn}$ between 46 and $89.5^\circ$ have been analyzed analogously to the above example. The ratio of the observed Maxwellian-temperature to the adiabatic-temperature (called "adiabaticity") is plotted in Figure~\ref{fig:adiabaticity} versus the Alfv\'en Mach number.  It is seen that, at individual crossings, thermal electrons may be heated more or less than adiabatically by factors as great as three, although the average adiabaticity is $1.06 \pm 0.52$.  The adiabaticity depends on the Alfv\'en Mach number, with the adiabaticity decreasing with increasing Mach number.  In this data set, there are no examples of sub-adiabatic heating for Mach numbers less than five and the few measurements above a Mach number of 10 all have adiabaticity less than one. The dashed curve in Figure 3 is proportional to $1/M_A$ and it describes the average behavior of the adiabaticity as a function of $M_A$, as is expected because $T_{e,\perp} \sim B$ and $M_A \sim 1/B$.

Parallel heating has also been studied by computing the Maxwellian-temperatures for electrons having pitch angles between $0^\circ-30^\circ$ and $150^\circ-180^\circ$.  The average [(parallel heating)/(perpendicular heating)] is $0.95\pm0.14$, which means that the parallel and perpendicular electrons are generally heated by the same amount when crossing the bow shock.
\begin{figure}[tbp]
\includegraphics[width=0.9\columnwidth]{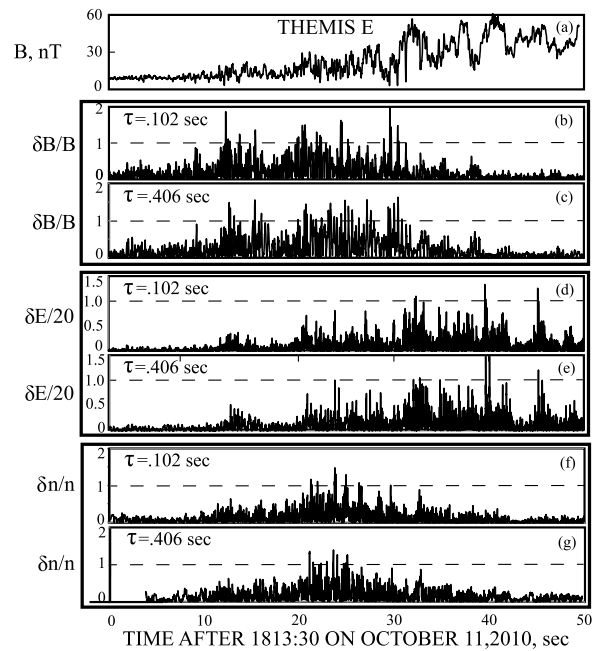}
\caption{Dimensionless perturbation ratios of magnetic field, electric field and density for time scales $\tau$ around the convected electron gyrodiameter. Quantities are measured with 128 samples/s. (a) Total magnetic field. (b-c) $\delta B/B$.  (d-e) $\delta E \cdot 2\rho_e/T_{e,\perp} \sim \delta E/20$. (f-g) $\delta n/n$.}
\label{fig:dBB}
\end{figure}

The conservation of the first adiabatic invariant, $T_{e,\perp}/B$, is derived under the assumption that the magnetic field does not change over the gyration orbit of an electron or the time for it to complete one gyration. Also assumed is that the electric potential across the electron gyrodiameter is small compared to the electron thermal energy, i.e. that the electric field gradient is small.  To test the validity of these conditions, we compute the fluctuations of the magnetic field, B, and electric field, E.
In this case the magnetic field requirement for conservation of the first invariant becomes
$\delta B/B(t,\tau) = [B(t+\tau)-B(t)]/[B(t+\tau)+B(t)]/2  \ll 1 $
where $\tau$ is a time scale, here corresponding to the convected gyrodiameter.

In Figure~\ref{fig:dBB}, $\delta B/B$ is calculated for a randomly selected bow shock crossing by THEMIS E, for which $\delta B$ was determined from search coil data and B came from fluxgate data. The local time of the crossing was 12:40 and the latitude was $11^\circ$, with the characteristic shock parameters $M_A=8.8$, $\theta_{Bn}=64^\circ$, and plasma $\beta =3.5$. 
Panel (a) of this figure gives the magnetic field strength and panels (b-c) give the time domain values of $\delta B/B$ for the time scales, $\tau$, associated with each panel.  For $T_{e,\perp}$ = 20 eV, B = 20 nT, and a shock velocity of 5 km/sec, the electron gyrodiameter is about 1 km and the time for the shock to move one gyrodiameter is about 0.2 seconds. Over a typical shock ramp the gyrodiameter may change by a factor of about two, so appropriate time scales, corresponding to the shock traveling one electron gyrodiameter, are chosen accordingly.
For these time scales there are substantial intervals in the ramp where $\delta B/B$ is the order of or larger than unity. The $\delta B/B$ is on average large for scales ranging from $c/\omega_{pe}$ to $\rho_i$ (not shown). These scales are spatial in the electron frame, when the thermal speed of these electrons is much greater than the phase speeds of whistler or lower-hybrid waves in this frequency regime. Finite amplitude fluctuations on shorter timescales (temporal) will also break adiabaticity. Thus, electrons cannot obey adiabatic conservation at such crossings.

A similar calculation may be done for the electric field fluctuations. Here the requirement is that the voltage across the electron gyrodiameter be small compared to the electron temperature, or
$\delta E \cdot 2\rho_e/T_{e,\perp} \ll 1 $
for the electrons to be adiabatic,
where $\rho_e$ is the electron gyroradius.  For $T_{e,\perp} \sim 20$~eV and $2\rho_e \sim 1$~km, $T_{e,\perp}/2\rho_e$ is 20~mV/m.  So this requirement becomes the condition that $\delta E/20$ be small for the electrons to have adiabatic trajectories. Figure~\ref{fig:dBB}, panels (d-e), show
$\delta E/20$ for the same time scales as for the magnetic field.
The electric field component in this calculation is the normal component found from a minimum variance analysis. The electric field fluctuations are comparable to or greater than one, so the electrons cannot move adiabatically in such a fluctuating electric field. It is interesting to note that the magnetic field fluctuations of Figure~\ref{fig:dBB} dominate near the foot of the ramp and the electric field fluctuations are most important near the top of the ramp. These fluctuations are not bipolar electrostatic structures \cite{bale2002b} at debye scales in the ramp. We also observe such structures but they are intermittently spaced, while the electron heating happens continuously throughout the ramp, which suggest that the electrons are scattered in the turbulent fields.  
We also note that the density fluctuations in Fig.~\ref{fig:dBB} (f-g) are large, $\delta n/n \sim 1$, so particle trapping and wave growth in density holes could be important.  

In conclusion, due to the large amplitude fluctuations of both B and E, the electron trajectories must be more random and chaotic than orderly and adiabatic. The thermal heating of electrons is therefore likely to be of stochastic nature. This result explains both the non-adiabatic heating of the perpendicular electrons and why the parallel and perpendicular heating are comparable. 

\begin{acknowledgments}
This work was supported by NASA contract NAS5-02099-09/12 and NASA grants NNX13AE24G, NNX09AE41G-1/14. The authors acknowledge the considerable assistance of Dr. J. P. McFadden and Professor S. Bale in the collection and interpretation of the plasma data.
\end{acknowledgments}

\bibliographystyle{abbrv}
% \bibliographystyle{plain}
% \bibliography{davidsbibliography}

\begin{thebibliography}{10}

\bibitem{angelopoulos2008a}
V.~{Angelopoulos}.
\newblock {The THEMIS Mission}.
\newblock {\em Space Science Reviews}, 141:5--34, Dec. 2008.

\bibitem{bale2002b}
S.~D. {Bale}, A.~{Hull}, D.~E. {Larson}, R.~P. {Lin}, L.~{Muschietti}, P.~J.
  {Kellogg}, K.~{Goetz}, and S.~J. {Monson}.
\newblock {Electrostatic Turbulence and Debye-Scale Structures Associated with
  Electron Thermalization at Collisionless Shocks}.
\newblock {\em ApJ}, 575:L25--L28, Aug. 2002.

\bibitem{balikhin1994a}
M.~{Balikhin} and M.~{Gedalin}.
\newblock {Kinematic mechanism of electron heating in shocks: Theory vs
  observations}.
\newblock {\em Geophys.~Res.~Lett.}, 21:841--844, May 1994.

\bibitem{balikhin1993a}
M.~{Balikhin}, M.~{Gedalin}, and A.~{Petrukovich}.
\newblock {New mechanism for electron heating in shocks}.
\newblock {\em Physical Review Letters}, 70:1259--1262, Mar. 1993.

\bibitem{balikhin1998a}
M.~{Balikhin}, V.~V. {Krasnosel'skikh}, L.~J.~C. {Woolliscroft}, and
  M.~{Gedalin}.
\newblock {A study of the dispersion of the electron distribution in the
  presence of E and B gradients: Application to electron heating at
  quasi-perpendicular shocks}.
\newblock {\em J.~Geophys.~Res.}, 103:2029--2040, Feb. 1998.

\bibitem{gedalin1995a}
M.~{Gedalin}, K.~{Gedalin}, M.~{Balikhin}, and V.~{Krasnosselskikh}.
\newblock {Demagnetization of electrons in the electromagnetic field structure,
  typical for quasi-perpendicular collisionless shock front}.
\newblock {\em J.~Geophys.~Res.}, 100:9481--9488, June 1995.

\bibitem{gedalin1995b}
M.~{Gedalin}, K.~{Gedalin}, M.~{Balikhin}, V.~{Krasnosselskikh}, and L.~J.~C.
  {Woolliscroft}.
\newblock {Demagnetization of electrons in inhomogeneous E{$\perp$}B:
  Implications for electron heating in shocks}.
\newblock {\em J.~Geophys.~Res.}, 100:19911--19918, Oct. 1995.

\bibitem{goodrich1984a}
C.~C. {Goodrich} and J.~D. {Scudder}.
\newblock {The adiabatic energy change of plasma electrons and the frame
  dependence of the cross-shock potential at collisionless magnetosonic shock
  waves}.
\newblock {\em J.~Geophys.~Res.}, 89:6654--6662, Aug. 1984.

\bibitem{kennel1985a}
C.~F. {Kennel}, J.~P. {Edmiston}, and T.~{Hada}.
\newblock {A quarter century of collisionless shock research}.
\newblock {\em Washington DC American Geophysical Union Geophysical Monograph
  Series}, 34:1--36, 1985.

\bibitem{lembege2003a}
B.~{Lemb{\`e}ge}, P.~{Savoini}, M.~{Balikhin}, S.~{Walker}, and
  V.~{Krasnoselskikh}.
\newblock {Demagnetization of transmitted electrons through a
  quasi-perpendicular collisionless shock}.
\newblock {\em Journal of Geophysical Research (Space Physics)}, 108:1256, June
  2003.

\bibitem{papadopoulos1985a}
K.~{Papadopoulos}.
\newblock {Microinstabilities and anomalous transport}.
\newblock {\em Washington DC American Geophysical Union Geophysical Monograph
  Series}, 34:59--90, 1985.

\bibitem{savoini2010a}
P.~{Savoini} and B.~{Lembege}.
\newblock {Non adiabatic electron behavior through a supercritical
  perpendicular collisionless shock: Impact of the shock front turbulence}.
\newblock {\em Journal of Geophysical Research (Space Physics)},
  115(A14):11103, Nov. 2010.

\bibitem{savoini2005a}
P.~{Savoini}, B.~{Lemb{\`e}ge}, V.~{Krasnosselskhik}, and Y.~{Kuramitsu}.
\newblock {Under and over-adiabatic electrons through a perpendicular
  collisionless shock: theory versus simulations}.
\newblock {\em Annales Geophysicae}, 23:3685--3698, Dec. 2005.

\bibitem{schwartz2011a}
S.~J. {Schwartz}, E.~{Henley}, J.~{Mitchell}, and V.~{Krasnoselskikh}.
\newblock {Electron Temperature Gradient Scale at Collisionless Shocks}.
\newblock {\em Physical Review Letters}, 107(21):215002, Nov. 2011.

\bibitem{schwartz1988a}
S.~J. {Schwartz}, M.~F. {Thomsen}, S.~J. {Bame}, and J.~{Stansberry}.
\newblock {Electron heating and the potential jump across fast mode shocks}.
\newblock {\em J.~Geophys.~Res.}, 93:12923--12931, Nov. 1988.

\bibitem{scudder1995a}
J.~D. {Scudder}.
\newblock {A review of the physics of electron heating at collisionless
  shocks}.
\newblock {\em Advances in Space Research}, 15:181--223, 1995.

\bibitem{scudder1986a}
J.~D. {Scudder}, T.~L. {Aggson}, A.~{Mangeney}, C.~{Lacombe}, and C.~C.
  {Harvey}.
\newblock {The resolved layer of a collisionless, high beta, supercritical,
  quasi-perpendicular shock wave. I - Rankine-Hugoniot geometry, currents, and
  stationarity}.
\newblock {\em J.~Geophys.~Res.}, 91:11019--11052, Oct. 1986.

\bibitem{scudder1986b}
J.~D. {Scudder}, T.~L. {Aggson}, A.~{Mangeney}, C.~{Lacombe}, and C.~C.
  {Harvey}.
\newblock {The resolved layer of a collisionless, high beta, supercritical,
  quasi-perpendicular shock wave. II - Dissipative fluid electrodynamics}.
\newblock {\em J.~Geophys.~Res.}, 91:11053--11073, Oct. 1986.

\bibitem{scudder1986c}
J.~D. {Scudder}, A.~{Mangeney}, C.~{Lacombe}, C.~C. {Harvey}, and C.~S. {Wu}.
\newblock {The resolved layer of a collisionless, high beta, supercritical,
  quasi-perpendicular shock wave. III - Vlasov electrodynamics}.
\newblock {\em J.~Geophys.~Res.}, 91:11075--11097, Oct. 1986.

\bibitem{sundkvist2012a}
D.~{Sundkvist}, V.~{Krasnoselskikh}, S.~D. {Bale}, S.~J. {Schwartz},
  J.~{Soucek}, and F.~{Mozer}.
\newblock {Dispersive Nature of High Mach Number Collisionless Plasma Shocks:
  Poynting Flux of Oblique Whistler Waves}.
\newblock {\em Phys.~Rev.~Lett.}, 108(2):025002, Jan. 2012.

\bibitem{treumann2009a}
R.~A. {Treumann}.
\newblock {Fundamentals of collisionless shocks for astrophysical application,
  1. Non-relativistic shocks}.
\newblock {\em A\&A~Rev.}, 17:409--535, Dec. 2009.

\end{thebibliography}

\end{document}